\documentclass[aps,pre,showpacs,twocolumn]{revtex4}
\usepackage{graphics}
\usepackage[dvips]{color}
\usepackage{graphicx}
\usepackage{amssymb}
\usepackage{rotating}
\usepackage{epsfig}

\input epsf.sty

\newcommand{\sts}{\scriptsize}

\newcommand{\mb}{\mbox}

\begin{document}

\title{Continuous phase transitions with a convex  dip in the
microcanonical entropy}
\author{Hans Behringer}
\affiliation{Fakult\"at f\"ur Physik, Universit\"at Bielefeld, D -- 33615
Bielefeld, Germany}
\author{Michel Pleimling}
\affiliation{Institut f\"ur Theoretische Physik I,
Universit\"at Erlangen-N\"urnberg, D -- 91058 Erlangen, Germany}

\begin{abstract}
  The appearance of a convex dip in the microcanonical entropy of
  finite systems usually signals a first order transition.
  However, a convex dip also shows up in some systems with a
  continuous transition as for example 
  in the Baxter-Wu model and in the four-state Potts model
  in two dimensions. We demonstrate that the appearance of a convex
  dip in those cases can be traced back to a finite-size effect. The
  properties of the dip are markedly different from those associated
  with a first order transition and can be understood within a
  microcanonical finite-size scaling theory for continuous
  phase transitions. Results obtained from numerical simulations corroborate
  the predictions of the scaling theory.

\end{abstract}

\maketitle



\section{Introduction}

The microcanonical analysis of phase transitions has gained growing
interest in recent years due to the inequivalence of the canonical and
the microcanonical descriptions of finite systems with short-range
interactions as well as of systems with long-range interactions
\cite{Gross_2001,Pleimling_2005,
  thirr70,bixon89,Hueller_1994,elli00,daux00,Gro00b,Plei01,Bar01,Isp01,gulm02,
  leyv02,bouch05,cost05,cost06,kast06}.  Whereas the discrimination of
the order of a phase transition from finite-size data can be a
difficult task in the canonical ensemble, microcanonical quantities
usually yield clear signatures which reveal the continuous or the
discontinuous character of a phase transition \cite{Pleimling_2005}.
In small microcanonical systems which exhibit a continuous transition
in the thermodynamic limit the character of the transition is signaled
by typical features of symmetry breaking, as e.\,g. the abrupt onset
of a non-zero order parameter when a pseudo-critical point is
approached from above. These features are accompanied by singular
physical quantities \cite{Promberger_Hueller_1995, Kastner_etal_2000,
  Hueller_Pleimling_2002, Behringer_2003, Behringer_2004, Hove_2004,
  Naudts_2004, Pleimling_etal_2004, Behringer_etal_2005,
  Richter_2005}. The associated singularities, however, are not the
consequence of a non-analytic microcanonical entropy and can therefore 
be characterized by classical critical exponents
\cite{Kastner_etal_2000, Behringer_2004}.  Similarly, intriguing
features are also revealed in the microcanonical entropy of small
systems with a discontinuous transition in the infinite volume limit
\cite{Hueller_1994,Brown_Yegulalp_1991, Schmidt_1994,
  Wales_Berry_1994, Gross_etal_1996, Deserno_1997, Ota_Ota_2000}. A
typical back-bending of the microcanonical caloric curve is observed,
leading to a negative heat capacity. This property, which has also
been observed in recent experiments \cite{Dag00,Sch01a}, has been used
as a tool for the determination of the first order character of phase
transitions in microcanonical systems.  In the canonical context the
first order character is then signaled by a double-peak structure in
certain histograms (e.\,g. \cite{Lee_1990, Lee_1991,Janke_1998}).

However, some systems (as for example the Baxter-Wu model 
\cite{Baxter_Wu_1973,Baxter_1982}),
with a continuous phase transition in
the thermodynamic limit, have been reported to show a similar
double-peak structure in canonical histograms or, equivalently, a negative
microcanonical specific heat \cite{Fukugita_1990,
Schreiber_2005}.  
It follows from this observation that 
the mere existence of a convex dip in the entropy is not
sufficient for the identification of a discontinuous transition
\cite{Stemmer_1998}.

In this work we examine more closely the finite-size behaviour of the
microcanonical specific heat and of the convex dip in the entropy
function of this kind of systems, thereby taking the Baxter-Wu model
and the four-state Potts model in two dimensions as examples. In
particular we demonstrate that the microcanonical finite-size scaling
theory presented in Reference \cite{Behringer_etal_2005} for
continuous phase transitions relates the appearance of these features
to a peculiar finite-size effect. In addition we deduce properties of
the dip in the microcanonical entropy which we test in numerical
simulations of both the Baxter-Wu and the four-state Potts model.

The rest of the article is organized as follows. In Section
\ref{sec:signaturen} a brief introduction to the microcanonical
analysis of physical systems is given. We particularly discuss the
properties of the microcanonical specific heat both for systems with
continuous and discontinuous phase transitions in the infinite volume
limit. The microcanonical finite-size
scaling theory of Reference \cite{Behringer_etal_2005} is used 
in Section \ref{sec:microfinite} to
deduce the scaling properties of the convex intruder in the
microcanonical specific heat of systems undergoing a continuous phase
transition in the thermodynamic limit.  In Section \ref{sec:potts} we
present numerical results both for the Baxter-Wu and for the
two-dimensional four-state Potts model which belong to the same
universality class. These numerical results corroborate the findings
obtained from the finite-size scaling theory as the properties of the
convex dip are consistent with the theoretical predictions.  Finally,
Section \ref{sec:ende} summarizes our results.

\section{Signatures of phase transitions in the microcanonical
  specific heat}
\label{sec:signaturen}

In finite systems the thermostatic quantities of a system undergoing a
phase transition in the thermodynamic limit show characteristic
features.  The canonical specific heat of finite systems, for example,
is a regular function that exhibits a maximum at a certain
temperature. The continuous or discontinuous character of the phase
transition is revealed by the size-dependence of the increase of the
specific heat maximum which is well understood within canonical finite-size scaling
theory \cite{Barber_1983, Privman_1990}.  In the case of a continuous
phase transition the increase of the specific heat maximum $c_{\sts \mb{max}}$ with the system
size $L$ is given by $c_{\sts \mb{max}} \sim L^{\alpha/\nu}$ where $\alpha$ describes
the divergence of the specific heat of the infinite system when
approaching the critical temperature (here and in the following we
only consider the case $\alpha > 0$), whereas $\nu$ is the usual
correlation length critical exponent.  For a discontinuous phase
transition, however, one finds $c_{\sts \mb{max}} \sim L^d$ \cite{Challa_1986} 
where $d$ is the
dimensionality of the system.

The thermodynamic behaviour can also be
investigated in the microcanonical ensemble which describes
systems which are isolated from any
environment \cite{Gross_2001,Pleimling_2005}.
The starting point in the microcanonical analysis is the
density of states $\Omega$ which, for a discrete spin system, measures 
the degeneracy of the macrostate $(E)$, i.e. $\Omega(E)$ is the number
of configurations (or microstates) with energy $E$. For a
$d$-dimensional system with linear extension $L$ physical quantities
are then calculated from the microcanonical entropy density
\begin{equation}
  s(e,L^{-1}) := \frac{1}{L^d}\ln \Omega(L^de, L^{-1}) 
\end{equation}   
which depends on the energy density $e := E/L^d$ and the system size
$L$. In the following the dependence on the system size is
indicated by the inverse system size $l:=L^{-1}$.
Note that the specific entropy also depends on the boundary
conditions which have to be specified for finite systems. This
dependence, however, is not indicated in
the notation here. The inverse microcanonical
temperature $\beta_\mu$ shows up as the conjugate variable to the natural
variable energy $e$ and is defined by 
\begin{equation}
  \beta_\mu(e,l) := \frac{\mb{d}}{\mb{d}e} s(e, l).
\end{equation}
The corresponding microcanonical specific heat is given by
\begin{equation}
\label{eq:spezallgfuermikro}
c(e, l) = -(\beta_\mu(e, l))^2\left(\frac{\mb{d}\beta_\mu(e, l)}{\mb{d}e}  \right)^{-1}.
\end{equation}

Similar to the canonical case the finite-size behaviour of the
microcanonical specific heat shows also certain characteristics if the
system has a phase transition in the infinite lattice. In the case of
a continuous transition the entropy density is concave everywhere. The
microcanonical specific heat (\ref{eq:spezallgfuermikro}) is thus
positive and displays a maximum whose value increases with growing
system size $L$ \cite{Behringer_etal_2005, Hove_2004}.  As the
derivative of $\beta_\mu(e,l)$ is the second order derivative of the
entropy, it measures the curvature of the entropy density. This
curvature is negative and has a maximum at a certain energy which can
be identified as the pseudo-critical energy $e_{\sts \mb{pc}}$ of the
finite system. The value of the derivative of $\beta_\mu$ at this
pseudo-critical energy scales like
\begin{equation}
\label{eq:spezskalen}
  \beta^\prime_\mu (e_{\sts \mb{pc}},l) := \frac{\mb{d}}{\mb{d}e}\beta_\mu(e,l)|_{e=e_{\sts \mb{pc}}} = 
  \frac{\mb{d}^2}{\mb{d}e^2}s(e,l)|_{e=e_{\sts \mb{pc}}} \sim l^{\alpha_\varepsilon/\nu_\varepsilon}
\end{equation}
in the regime $l\to 0$. The microcanonical critical exponent
$\alpha_\varepsilon$ is thereby related to the canonical critical
exponent $\alpha$ by $\alpha_\varepsilon = \alpha/(1-\alpha)$
\cite{Promberger_Hueller_1995,Kastner_etal_2000}.  Generally, a
canonical critical exponent $\kappa$ translates to the microcanonical
exponent $\kappa_\varepsilon = \kappa/(1-\alpha)$ if $\alpha$ is
positive, whereas for non-positive $\alpha$ the critical exponents of
the microcanonical description are identical to the canonical ones
\cite{Behringer_2005}.  Relation (\ref{eq:spezskalen}) is obtained
from a decomposition of the microcanonical entropy into a regular part
and a singular part which obeys a certain homogeneity relation in the
vicinity of the pseudo-critical point. Note that the size-dependence
of the increase of the specific heat maximum allows the determination
of the corresponding microcanonical critical exponent of the specific
heat of the infinite system. For more details see Section
\ref{sec:microfinite} and in particular Reference
\cite{Behringer_etal_2005}.  Note also that other microcanonical quantities
like the spontaneous magnetization and the susceptibility also exhibit
typical features of continuous phase transitions in finite systems
\cite{Kastner_etal_2000, Hueller_Pleimling_2002, Behringer_2003,
  Behringer_2004, Pleimling_etal_2004, Richter_2005}.  These aspects
are however not studied in the present work.

\begin{figure}[h!]
\begin{center}
\epsfig{file=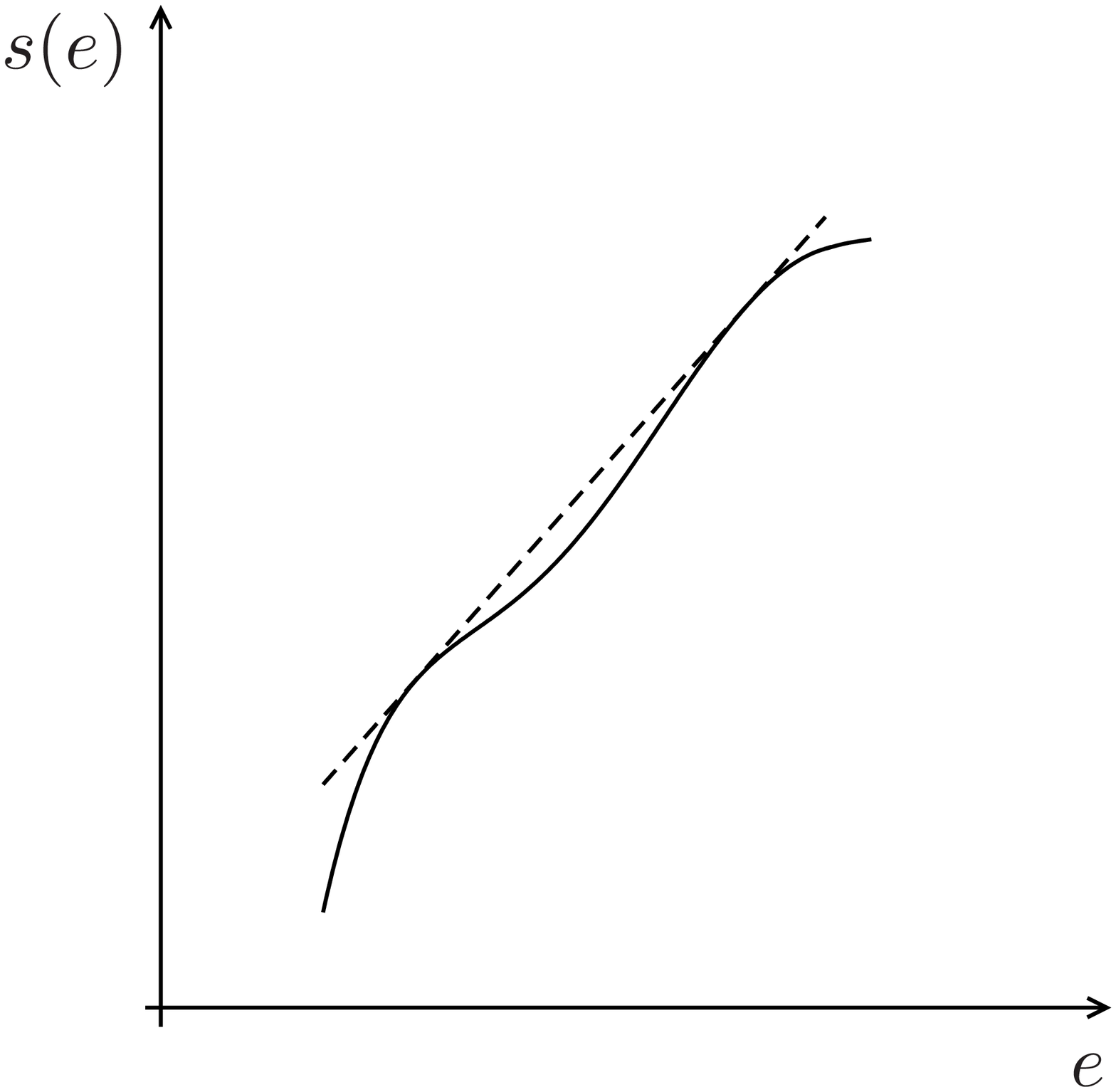,width=14.6em, angle=270}
\epsfig{file=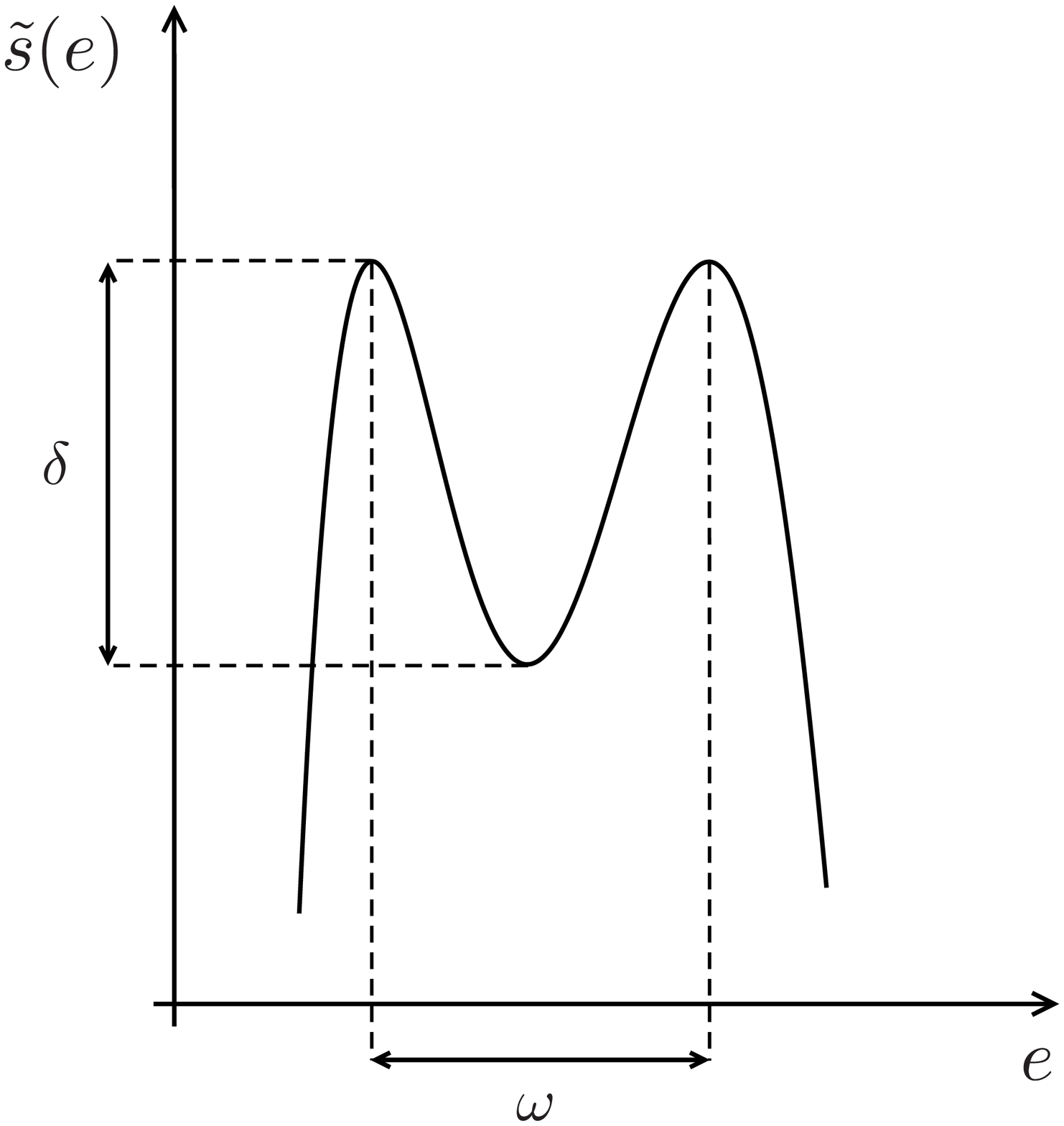,width=15em, angle=270}
\caption[Schematic depiction of the derivatives of the reduced
entropy]{\label{bild:schema_maxwelldip}\small Schematic depiction of
  the convex dip in the entropy of a finite system whose corresponding
  infinite system undergoes a
  discontinuous phase transition. The double-tangent with slope
  $\beta_{\sts \mb{t}}$ is indicated by a dashed line (top). For the 
  function $\tilde{s} = s - \beta_{\sts \mb{t}}e$ the definitions of the width
  $\omega$ and of the depth $\delta$ of the dip are shown (bottom).}
\end{center}
\end{figure}

\begin{figure}[h!]
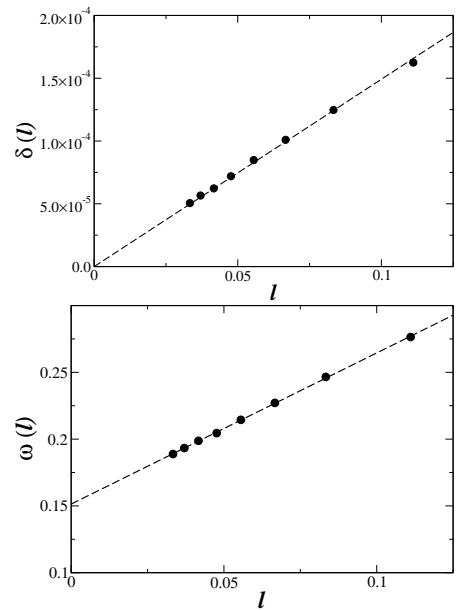

\begin{center}
\epsfig{file=figure2a.eps,width=18em, angle=0}
\epsfig{file=figure2b.eps,width=18em, angle=0}
\caption[Schematic depiction of the derivatives of the reduced
entropy]{\label{bild:3q3d}\small The depth $\delta(l)$ and width $\omega(l)$ of the convex dip
in the microcanonical entropy of the three-state Potts model in three
dimensions as a function of the inverse system size. The dashed lines are linear fits to the data.
Error bars are comparable to the size of the symbols.}
\end{center}
\end{figure}

The behaviour of the specific heat 
is completely different if the system undergoes a
discontinuous phase transition. In this case the microcanonical
entropy density of a finite system has a convex dip which leads to a
negative microcanonical specific
heat for a certain energy interval \cite{Brown_Yegulalp_1991,
  Hueller_1994, Schmidt_1994, Wales_Berry_1994, Gross_etal_1996,
  Deserno_1997, Ota_Ota_2000}. This is different in the canonical
ensemble where the specific heat,
which is related to the variance of the energy,
is always positive for any finite system.
Of course in the thermodynamic limit the convex dip in the
microcanonical entropy density disappears in systems with finite-range
interactions and the specific heat is positive everywhere.  The
curvature properties of the entropy density of a finite system allow
the determination of an inverse pseudo-transition temperature
$\beta_{\sts \mb{t}}$ by the double-tangent (or Maxwell) construction.
This is depicted schematically in Figure \ref{bild:schema_maxwelldip}.
The discontinuous character of the transition is revealed by a certain
scaling behaviour of the convex dip. Definig the auxiliary function
$\tilde{s}(e,l) := s(e,l) - \beta_{\sts \mb{t}}(l)e$, a size-dependent
width $\omega(l)$ and depth $\delta(l)$ of the dip can be measured, as
shown in Figure \ref{bild:schema_maxwelldip}. With increasing system
size, i.\,e. for the regime $l\to 0$, the depth scales like $\delta(l)
\sim l$ and the width like $\omega(l) - \omega_\infty \sim l$
\cite{Borgs_1990,Borgs_1992,Borgs_1991}. Here $\omega_\infty$ is the
length of the linear section of the function $\tilde{s}$ of the
infinite system and is in fact the latent heat of the transition. The
asymptotic behaviour linear in $l$ and the appearance of a non-zero
latent heat $\omega_\infty$ signal the discontinuous character of the
transition in the infinite system. This can be illustrated
by investigating the three-state Potts model in three dimensions. The
general $q$-state Potts model is defined by the Hamiltonian 
\begin{equation}
  \mathcal{H}(\sigma) = - \sum_{\left<ij\right>} \delta_{\sigma_i,\sigma_j}
\end{equation} 
where the sum runs over nearest neighbour sites of the lattice and the
spins $\sigma_i$ can take on the values $\sigma_i= 1,\ldots,q$  \cite{Potts_1952, Wu_1982,
  Baxter_1982}. The
coupling constant is set to unity. The three-state model in three dimensions undergoes a
(relatively weak) first order transition and therefore one expects
that the characteristic quantities of the convex dip evolve linearly
in the inverse system size (compare \cite{Janke_1997} for a canonical
and \cite{Schmidt_1994} for a microcanonical investigation of the model). This is indeed confirmed by numerical
data as shown in Figure \ref{bild:3q3d}.

\section{Microcanonical finite-size scaling theory}
\label{sec:microfinite}

In this Section the implications of the microcanonical finite-size
scaling theory \cite{Behringer_etal_2005}
are investigated for systems with a continuous phase
transition where $\alpha_\varepsilon \geq \nu_\varepsilon$
(canonically $\alpha \geq \nu$). 
Note that the condition $\alpha_\varepsilon\geq\nu_\varepsilon$
corresponds to a rather restricted class of model systems. Nevertheless
important spin models like the Baxter-Wu and the four-state Potts
model in two dimensions have  critical exponents 
$\alpha_\varepsilon = \nu_\varepsilon$ and therefore belong to this
class. These models are closer studied in the next Section. Other models are
reported in the literature to have exponents satisfying
$\alpha_\varepsilon > \nu_\varepsilon$ \cite{Badke_1985a,Badke_1985b}.

Consider the entropy density $s(e,l)$ of a finite $d$-dimensional
microcanonical system near the pseudo-critical energy $e_{\sts
  \mb{pc}}$. Recall that $e_{\sts \mb{pc}}$ is defined by the maximum
of the second derivative of $s$. The microcanonical finite-size
scaling theory for a system with a continuous transition 
\cite{Behringer_etal_2005} assumes that
the entropy can be split up into two parts, namely into a so-called
regular part $s_{\sts \mb{r}}$ and into a part $s_{\sts \mb{s}}$ 
which is called singular 
as it becomes non-analytic in the infinite volume limit. This is in contrast
to the regular part which is analytic also in the infinite system.
The singular part has to satisfy the scaling
relation
\begin{equation}
\label{eq:s_homogen}
s_{\sts \mb{s}}(\varepsilon,l) = l^d \phi (l^{-1/\nu_\varepsilon}\varepsilon)
\end{equation}
with $\varepsilon := e - e_{\sts \mb{pc}}$ and $\phi$ being a scaling
function with no further explicit size-dependence.
This decomposition $s(\varepsilon,l) = s_{\sts
  \mb{r}}(\varepsilon,l) + s_{\sts \mb{s}}(\varepsilon,l)$ is only
valid for asymptotically large system sizes and in the regime
$l^{-1/\nu_\varepsilon}\varepsilon \to 0$, in particular for energies
close to the pseudo-critical one. Note that the singular part and
thus the function $\phi$ in relation (\ref{eq:s_homogen}) is also
analytic for finite systems and must fulfil the required scaling
property which states that it is a homogeneous function. Due to the
analyticity of the microcanonical entropy the decomposition together
with the required scaling relation for the singular part states that
the entropy is of the form
\begin{equation}
\label{eq:ganzesred_entwickelt_vier}
    s(\varepsilon,l) = s_0(l) + \beta_{\sts \mb{pc}}(l)\varepsilon + \frac{1}{2}A_2(l)\varepsilon^2 +
    \frac{1}{4}A_4(l)\varepsilon^4 +\ldots
\end{equation}
for small $\varepsilon$ \cite{Behringer_etal_2005}.
Here $\beta_{\sts \mb{pc}}(l) = \beta_\mu(e_{\sts \mb{pc}},l)$.
The coefficients of the second and fourth order terms are of the form
\begin{equation}
\label{eq:a2_coef}
    A_2(l) = B_2l^{\frac{\alpha_{\varepsilon}}{\nu_{\varepsilon}}}
    + C_2(l)
\end{equation}
and
\begin{equation}
\label{eq:a4_coef}
A_4(l) = B_4l^{\frac{\alpha_{\varepsilon}-2}{\nu_{\varepsilon}}}
    + C_4(l)
\end{equation}
in terms of the critical exponents of the microcanonical system. The
coefficient $A_4$ has to be negative, otherwise higher order terms in
the energy deviation $\varepsilon$ have to be taken into account. The
size-dependence contained in the coefficients $C_2$ and $C_4$ stems
from the regular part $s_{\sts \mb{r}}$ of the entropy density whereas
the dependencies involving the critical exponents $\alpha_\varepsilon$
and $\nu_\varepsilon$ have their origin in the singular part $s_{\sts
  \mb{s}}$ (or equivalently in the scaling function $\phi$ in relation
(\ref{eq:s_homogen})) of the entropy. In particular, the coefficient
$B_2$ has to be negative in the case of a continuous transition and,
due to the regularity of $s_{\sts \mb{r}}$, the coefficient $C_2$ is of
the form 
\begin{equation}
\label{eq:C2}
C_2(l) = v_1l+v_2l^2 + \ldots
\end{equation}
for small $l$ whereas the
coefficient $C_4$ has the expansion $C_4(l) = z_0 + z_1l + \ldots$
in the inverse system size (see reference \cite{Behringer_etal_2005}
for further details).  Note that $\alpha_\varepsilon$ has to satisfy
$\alpha_\varepsilon \leq 2$ for $C_4$ to be the subdominant term in
the coefficient $A_4$. This is assumed in the following
considerations.

In certain model systems the critical exponents satisfy the relation
$\alpha_\varepsilon = \nu_\varepsilon$. Examples include the Baxter-Wu model and the
four-state Potts model in two dimensions. In this situation the
leading behaviour of the coefficient $A_2(l)$ in relation
(\ref{eq:ganzesred_entwickelt_vier}) is the linear term in the inverse
system size, i.\,e. $A_2(l) \sim a\; l$ for $l\to 0$ where the amplitude $a$ is the sum
$B_2 + v_1$. If the amplitude $a$ is positive, i.\,e. $v_1$ is
positive and large enough compared to the modulus of the negative
amplitude $B_2$, the entropy density develops
a convex dip for all finite system sizes. Note that the corresponding
condition $A_2(l)>0$ is a necessary condition for the appearance of a
convex dip in the microcanonical entropy. Thus, in the framework of the
microcanonical finite-size scaling theory a convex dip
can appear even so the transition in the infinite system is continuous.
Furthermore, from  expression
(\ref{eq:ganzesred_entwickelt_vier}) one can then deduce scaling relations
for the width and the depth of the convex dip in such a case. Applying
the double-tangent construction to the entropy
(\ref{eq:ganzesred_entwickelt_vier}) with $\alpha_\varepsilon =
\nu_\varepsilon$ and  a positive $A_2$ one finds
that
\begin{equation}
\label{eq:skalomega}
  \omega(l) = 2\sqrt{\frac{A_2(l)}{|A_4(l)|}} \stackrel{l\to 0}{\sim} l^{1/\alpha_\varepsilon} 
\end{equation}
  and
\begin{equation}
\label{eq:skaldelta}
  \delta(l) = \frac{(A_2(l))^2}{|A_4(l)|}\stackrel{l\to 0}{\sim} l^{1+2/\alpha_\varepsilon}
\end{equation}
for small $l$. This scaling behaviour of the convex dip in the
microcanonical entropy is therefore different from the linear
behaviour in the inverse system size encountered for discontinuous
transitions and discussed in Section \ref{sec:signaturen} \cite{Badke_1985a,Badke_1985b}. 
For a negative amplitude $a$, on the other hand, the signatures of
a continuous transition with the scaling behaviour $
\beta_\mu^\prime(e_{\sts \mb{pc}},l) \sim A_2(l) \sim l$ 
($ = l^{\alpha_\varepsilon/\nu_\varepsilon}$) are
observed. 
Whether the predicted convex dip really appears or whether one encounters
typical features of continuous transitions thus depends on the sign and
relative size of the amplitudes entering the scaling form
(\ref{eq:ganzesred_entwickelt_vier}) of the entropy near the
pseudo-critical energy. 


Let us end this Section by considering the
situation where
$\alpha_\varepsilon$ is  strictly larger than  $
\nu_\varepsilon$. In this case the leading behaviour of the
coefficient $A_2$ is still linear in $l$ provided the amplitude $v_1$
is non-zero. For positive $v_1$ one then has again a convex dip
with the scaling relations
$\omega(l) \sim l^{1/2 - (\alpha_\varepsilon-2)/(2\nu_\varepsilon)}$
and $\delta(l) \sim l^{2 - (\alpha_\varepsilon-2)/\nu_\varepsilon}$
for its characteristic extensions.
For negative $v_1$ instead one observes features of a continuous
transition with the asymptotic behaviour $
\beta_\mu^\prime(e_{\sts \mb{pc}},l) \sim l$,  which does not allow to
uncover the actual critical exponents from the leading size-dependence
of the evolution of the microcanonical specific heat, in contrast
to the corresponding situation for $\alpha_\varepsilon \leq
\nu_\varepsilon$. 

In conclusion, a convex dip in the microcanonical entropy is in general possible for
systems with a continuous phase transition if the critical exponents satisfy
$\alpha_\varepsilon \geq \nu_\varepsilon$. For systems with
$\alpha_\varepsilon > \nu_\varepsilon$ such a dip shows up if the
amplitude $v_1$ of the linear term in the expansion of $C_2$ in
(\ref{eq:C2}) is positive. In a situation with $\alpha_\varepsilon
= \nu_\varepsilon$ the amplitude $v_1$ has to satisfy the additional
requirement that $v_1 > |B_2|$. In both cases we end up with the necessary
condition $A_2(l) > 0$, see Eq. (\ref{eq:a2_coef}).

\section{Microcanonical specific heat of the Baxter-Wu and of the four-state Potts model}
\label{sec:potts}

\subsection{The Baxter-Wu model}
The Baxter-Wu model is a classical spin model whose free energy can be
calculated exactly in zero magnetic field \cite{Baxter_Wu_1973,
  Baxter_1982}. The model is defined on a two-dimensional triangular
lattice, with the Hamiltonian
\begin{equation}
\label{eq:baxterwudef}
  \mathcal{H}(\sigma) =- \sum_{\left< ijk \right>} \sigma_i\sigma_j\sigma_k
\end{equation}
where the strength of the coupling constant has been set to unity.
The spins $\sigma_i$ are Ising spins with the possible values
$\sigma_i = \pm 1$. The sum in (\ref{eq:baxterwudef}) extends over all
triangles of the triangular lattice (here denoted by $\left< ijk
\right>$ with $i,j$ and $k$ specifying the vertices of the triangles).
The system has a continuous phase transition in the infinite system at
the critical temperature $T_{\sts \mb{c}} = 2/\ln(1+\sqrt{2})$. The
critical exponents which characterize the singular behaviour of
physical quantities near the critical point are known exactly. For the
specific heat and the correlation length one has for example $\alpha =
\nu = 2/3$ or $\alpha_\varepsilon = \nu_\varepsilon = 2$
microcanonically. In this Section the microcanonical 
specific heat of the Baxter-Wu
model is investigated numerically \cite{martinos}. To this end the
microcanonical entropies of systems with linear sizes ranging from 16
to 96 are calculated using the very efficient transition observable
method \cite{Hueller_Pleimling_2002, Richter_2005} for both open and
periodic boundary conditions.

\begin{figure}[h!]
\begin{center}
\epsfig{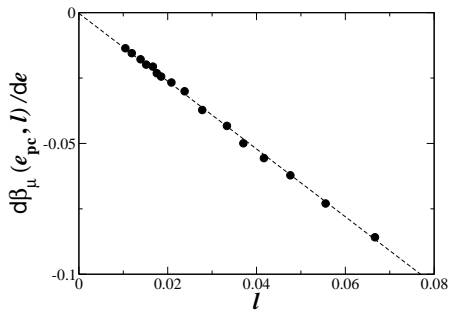}
\caption[Schematic depiction of the derivatives of the reduced
entropy]{\label{bild:baxwuoffen}\small The derivative
  $\beta_\mu^\prime(e_{\sts\mb{pc}},l) = \mb{d} \beta_\mu(e_{\sts\mb{pc}},l)/\mb{d}e$ of the Baxter-Wu model
  as a function of the inverse system size
  $l$ for open boundary conditions. The errors are approximately
  as large as the symbol size. The dashed line shows a linear fit.
  The  data extrapolate to zero
  in the limit $l \to 0$, corresponding to a divergent specific heat
  in the infinite system.}
\end{center}
\end{figure}

As the Baxter-Wu model exhibits a continuous phase transition in the
infinite lattice, one expects at first sight that the
microcanonical specific heat shows a maximum with the associated 
increase for increasing system sizes.
For finite systems with {\em open} boundary
conditions this is indeed the case. The derivative $
\beta_\mu^\prime(e_{\sts\mb{pc}},l) $ of the microcanonical temperature
at its maximum value, i.\,e. at the pseudo-critical energy, is negative
and its modulus decreases, leading to an increasing specific heat
maximum when
the system size is increased. For the Baxter-Wu model the critical
exponents $\alpha_\varepsilon$ and $\nu_\varepsilon$ are equal and
therefore one expects from (\ref{eq:spezskalen}) and the discussion in
Section \ref{sec:microfinite} to observe the scaling relation $
\beta_\mu^\prime(e_{\sts \mb{pc}},l) \sim l$ for small $l$. This
expectation is indeed confirmed by the data shown in Figure
\ref{bild:baxwuoffen}.

\begin{figure}[h!]
\begin{center}
\epsfig{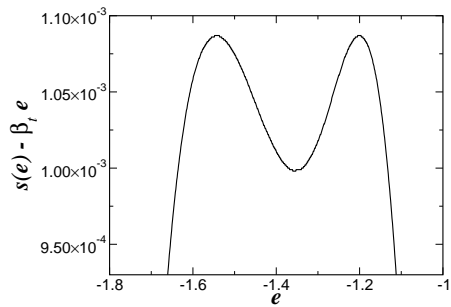}
\caption[Schematic depiction of the derivatives of the reduced
entropy]{\label{bild:baxwuperiodisch_dip}\small The function
  $\tilde{s}$ for the finite Baxter-Wu system with $L=60$ and periodic
  boundary conditions (here $\beta_{\sts \mb{t}} = 0.43945$). Note
  that $\tilde{s}$ is only obtained from a numerical simulation up to
  an additive constant. The data show a double-peak structure typical
  of finite systems with a discontinuous transition in the macroscopic
  limit.}
\end{center}
\end{figure}

The situation is different for finite Baxter-Wu systems with {\em
  periodic} boundary conditions. In Figure
\ref{bild:baxwuperiodisch_dip} the function $\tilde{s}$ 
is shown for a finite system
with linear extension $L= 60$. A convex dip is clearly visible. 
This observation
apparently suggests at first sight that the model seems to undergo a
{\em discontinuous} transition in contradiction to the known {\em
  continuous} character of the phase transition.
Note that recently
a double-peak structure in canonical histograms of the energy was
reported in the literature \cite{Schreiber_2005}.

\begin{figure}[h!]
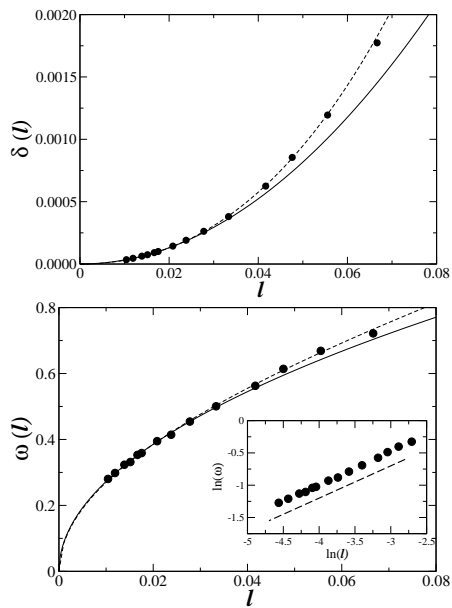

\begin{center}
\epsfig{file=figure5a.eps,width=18em, angle=0}
\epsfig{file=figure5b.eps,width=18em, angle=0}
\caption[Schematic depiction of the derivatives of the reduced
entropy]{\label{bild:baxwuperiodisch_dipsKalen}\small The depth
  $\delta(l)$ and width $\omega(l)$ as functions of the inverse system size
  $l$ for Baxter-Wu systems with periodic boundary conditions. The
  errors are of the order of the symbol size. The full lines
  show fits to the theoretically predicted behaviour  (\ref{eq:skalomega}) and
  (\ref{eq:skaldelta}) using the six largest system sizes.
  The dashed lines include in addition the first correction term
coming from the regular part $s_{\sts \mb{r}}$ of
the microcanonical entropy. The inset shows a double-logarithmic plot
  of the width together with a straight line of slope 1/2. From the
  six largest systems one gets a slope of 0.481(3). 
}
\end{center}
\end{figure}

In the previous Section \ref{sec:microfinite} we have shown that the
microcanonical finite-size scaling theory permits the appearance of
such a convex dip for the case $\alpha_\varepsilon = \nu_\varepsilon$.
With the relations (\ref{eq:skaldelta}) and (\ref{eq:skalomega}) and
the known value $\alpha_\varepsilon = 2$ for the Baxter-Wu model we
expect to observe the scaling behaviour $\omega(l) \sim \sqrt{l}$ for
the width and $\delta(l) \sim l^2$ for the depth of the convex dip.
These predictions are indeed confirmed by our numerical data obtained
for the Baxter-Wu model as shown in Figure
\ref{bild:baxwuperiodisch_dipsKalen}. The full lines are fitted
functions containing the expected asymptotic behaviour in the inverse
system size $l$. For the dashed lines the leading correction term
coming from the regular part $s_{\sts \mb{r}}$ of the microcanonical
entropy (which is proportional to $l^3$ for the depth and to $l^{3/2}$
for the width) has been included.  In agreement with the continuous
character of the phase transition the dip scales away in the
macroscopic limit $l\to 0$.  The size-dependences of both the depth
and the width of the convex dip follow the theoretical predictions.
This is especially clear from the inset of Figure
\ref{bild:baxwuperiodisch_dipsKalen} where we compare $\omega(l)$ with
the expected behaviour in a double-logarithmic plot. This agreement
gives strong evidence to the general scaling theory presented in the
previous Section \ref{sec:microfinite}.

It should be stressed again that the different behaviour of the
microcanonical entropy of the Baxter-Wu model for different boundary
conditions is related to different properties of the expansion
coefficients in ($\ref{eq:ganzesred_entwickelt_vier}$). In particular
the coefficient $A_2$ seems to be deeply affected by the boundary
conditions. A microscopic explanation of this
observation is however still lacking.

The findings of this section have direct implications for the
behaviour of canonical energy histograms, which have been discussed
for the Baxter-Wu model in Reference \cite{Schreiber_2005}. The energy
histograms are directly related to the canonical distribution function
$P_\beta(e)$ which is the probability of finding a microstate with
energy density $e$ at the inverse canonical temperature $\beta$. In terms of
the microcanonical entropy this distribution function is basically
given by $P_\beta(e) \sim \exp\left(\beta L^de - L^ds(e,l)\right)$.
Thus for the inverse pseudo-transition temperature $\beta_{\mb{\sts t}}$
obtained from the Maxwell construction applied to the microcanonical
entropy the canonical energy histogram exhibits two peaks at equal
height with the peaks appearing at the same energies as the peaks in
$\tilde{s} = s - \beta_{\mb{\sts t}}e $. It follows that in the canonical histograms the
peaks at the energies $e_1$ and $e_2$ approach
each other according to the scaling law $|e_1 - e_2| \sim \sqrt{l}$.

An important quantity which influences the kinetics of first order
phase transitions is the interface tension $\Sigma$ defined by
\begin{equation}
2\beta_{\mb{\sts t}}\Sigma = \frac{1}{L^{d-1}}
\ln\frac{P_{\beta_{\mb{\sts t}}}(e_{\mb{\sts
      max}})}{P_{\beta_{\mb{\sts t}}}(e_{\mb{\sts min}})}
\end{equation}
where the double-peaked distribution $P_{\beta_{\mb{\sts t}}}$ has one
of the two peaks of equal height at energy $e_{\mb{\sts max}}$
whereas the minimum between the two peaks appears at the energy
$e_{\mb{\sts min}}$ \cite{Binder_1987}.  From the above discussion one
gets $\Sigma \sim \delta(l)/l\sim l$ and thus the interface tension
vanishes in the limit of asymptotically large systems. Consequently a
coexistence between an ordered phase and a disordered phase at
temperature $\beta_{\mb{\sts t}}$ is hardly detectable in this case
although the canonical distribution function $P_{\beta_{\mb{\sts t}}}$
exhibits a double-peak structure. For moderately small systems,
however, the dip represents still a considerable barrier between the
order phase and the disordered one. In some sense one can speak of a
phase coexistence for not too large finite systems.

\subsection{The four-state Potts model}
The four-state Potts model in two dimensions belongs to the same
universality class as the Baxter-Wu model. 
The character of the phase transition of the $q$-state
Potts model in two dimensions changes exactly at $q=4$ from continuous for low values of $q$ to
discontinuous for large $q$.
This makes the four-state
Potts model particularly interesting. Due to the presence of marginal
scaling fields for $q=4$ the singular behaviour of the four-state
Potts model is not simply described by pure power laws, but physical
quantities like the specific heat or the susceptibility acquire additional logarithmic
factors \cite{Nauenberg_1980,Salas_1997}. 
Logarithmic corrections then also show up in the 
finite-size scaling form of the specific heat
\cite{Barber_1983}. This is in contrast to the Baxter-Wu model where these
marginality effects are absent and the singular behaviour of the
specific heat does not exhibit logarithmic corrections
\cite{Nauenberg_1980,Novotny_1982}.

\begin{figure}[h!]
\begin{center}
\epsfig{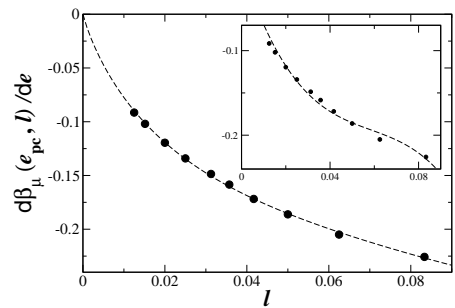}
\caption[Schematic depiction of the derivatives of the reduced
entropy]{\label{bild:pottsoffen}\small The derivative
  $\beta_\mu^\prime(e_{\sts\mb{pc}},l) = \mb{d} \beta_\mu(e_{\sts\mb{pc}},l)/\mb{d}e$ of the four-state Potts
  model as a function of the inverse system size
  $l$ for open boundary conditions. The errors are approximately
  of the order of the symbol size. The dashed line shows a fit with an
  asymptotic linear term and corrections including a logarithmic term.
  The  data extrapolate to zero
  for the limit $l \to 0$, yielding a divergent specific heat
  in the infinite system. The inset shows a fit without
  the logarithmic correction term (see main text for further details).}
\end{center}
\end{figure}

Using the same numerical technique as for the Baxter-Wu model, we have
computed the microcanonical entropy density of the four-state
Potts model on
a square lattice with open and periodic boundary conditions.
For open boundary conditions system sizes ranging from
$L=12$ to $L=80$ have been considered, whereas for periodic boundary conditions
system sizes from $L=14$ to $L=50$ have been simulated. 

As shown in Figure
\ref{bild:pottsoffen} the derivative $
\beta_\mu^\prime(e_{\sts\mb{pc}},l) $ of the microcanonical temperature
evaluated at the pseudo-critical energy is found to be negative 
for open boundary conditions.  
As its modulus decreases as a function of the inverse system size,
the resulting specific heat maximum increases when the
system size is increased, thus yielding the typical finite-size features of a
continuous transition in the thermodynamic limit. The
critical exponents $\alpha_\varepsilon$ and $\nu_\varepsilon$ being
equal for the four-state Potts model, one expects
(see Eq. (\ref{eq:spezskalen})) the scaling relation $
\beta_\mu^\prime(e_{\sts \mb{pc}},l) \sim l$ for small $l$. However, as
Figure \ref{bild:pottsoffen} demonstrates, this asymptotic regime is
not yet fully reached for the system sizes considered here. Presumably, this has
its origin in the presence of logarithmic correction terms.
Nevertheless, the data are consistent with
the expectation of a leading linear dependence in the inverse system
size in the asymptotic limit $l\to 0$. To demonstrate this a function
of the form $(\beta_\mu^\prime(e_{\sts\mb{pc}},l))_{\sts \mb{fit}}
= al + bl/\ln(l) +cl^2$ has been fitted to the data (dashed line in
Figure \ref{bild:pottsoffen}).  To get a rough impression of the
presence of a logarithmic correction, a fit function with correction
terms $b_1l^2 + c_1l^3$, i.\,e. without a logarithmic term but with the
same number of fit parameters, has also been fitted to the numerical data,
yielding a much poorer result (see inset of Figure \ref{bild:pottsoffen}).

\begin{figure}[h!]
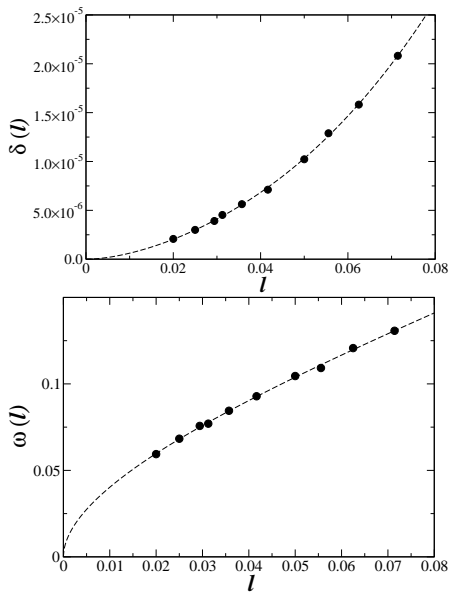

\begin{center}
\epsfig{file=figure7a.eps,width=18em, angle=0}
\epsfig{file=figure7b.eps,width=18em, angle=0}
\caption[Schematic depiction of the derivatives of the reduced
entropy]{\label{bild:pottsperiodisch_dipsKalen}\small The depth
  $\delta(l)$ and width $\omega(l)$ as functions of the inverse system size
  $l$ for four-state Potts systems with periodic boundary conditions. The
  errors are approximately as large as the symbol size. The dashed lines
  show fits of the expected scaling relations (\ref{eq:skalomega}) and
  (\ref{eq:skaldelta}) to the data  with a logarithmic correction term.
}
\end{center}
\end{figure}

For the Potts model with periodic boundary conditions our findings are
in qualitative agreement with the considerations of
Section \ref{sec:microfinite} and with the results obtained for the
Baxter-Wu model with periodic boundaries. The microcanonical entropy
displays the same convex dip for finite systems.  Note that a
double-peak structure in energy histograms obtained from canonical
Monte Carlo simulations was already reported in the literature for
this model \cite{Fukugita_1990}.  The scaling behaviours of the
depth and the width of this dip are again consistent with the
predicted scaling relations (\ref{eq:skalomega}) and
(\ref{eq:skaldelta}) as depicted in Figure
\ref{bild:pottsperiodisch_dipsKalen}, even so we were forced to
include a logarithmic correction term in addition to the leading
correction term coming from the regular part.

\section{Conclusions}
\label{sec:ende}

The determination of the order of a phase transition presents a basic
issue in the statistical analysis of physical systems. As there are
only few model systems that can be tackled analytically this
question is usually addressed through numerical simulations. A widely
used indicator of a first order transition is the metastability
between two coexisting phases \cite{Fukugita_1990, Lee_1990,
  Lee_1991}. This leads to a convex dip in the microcanonical entropy
and thus to a negative microcanonical specific heat in the transition region. In the
canonical ensemble the convex dip corresponds to a double-peak
structure in histograms of suitable observables. However, the appearance of a
double-peak structure in canonical histograms or a convex dip in the
microcanonical entropy is not sufficient to identify a first order
transition as such features have been reported in the
literature for certain model systems that are known to undergo a
continuous transition in the thermodynamic limit. To identify the
first order character additional scaling properties of the convex dip
in the microcanonical entropy (or the double-peak structure in
canonical histograms \cite{Lee_1990, Lee_1991}) have to be satisfied.

In this work we analyzed the Baxter-Wu model and the two-dimensional
four-state Potts model, which are known to have continuous transitions
in the thermodynamic limit, but for which double-peak structures in
canonical histograms are observed \cite{Fukugita_1990,Schreiber_2005}.
Using a recently developed phenomenological finite-size scaling theory
for systems with continuous phase transitions
\cite{Behringer_etal_2005} the possible properties of the
microcanonical entropy of models where $\alpha \geq \nu$ have been
investigated. It turns out that finite-size effects may indeed lead to
a convex dip in the entropy of any finite system although the models
show continuous transitions. The finite-size scaling relations for the
microcanonical entropy predict in this case certain scaling properties
of the convex dip that are different from those encountered at a first
order transition. The numerical studies of the Baxter-Wu system and
the four-state Potts model with periodic boundary conditions indeed
reveal an entropy with a convex dip. The properties of the convex dip
are in agreement with the expected scaling relations deduced from the
microcanonical finite-size scaling theory.  This gives strong support
to the validity of the scaling theory proposed in
\cite{Behringer_etal_2005} for finite microcanonical systems. The
systems with open boundaries on the other hand do not show a convex
dip in the corresponding entropy. For this case too, the computed
specific heat satisfies the scaling relation of the microcanonical
scaling theory. The different behaviour for different boundary
conditions could be traced back to the qualitatively different
properties of the expansion coefficient $A_2$ in
(\ref{eq:ganzesred_entwickelt_vier}). Future studies have to clarify
in more details the influence of different boundary conditions on the
finite-size behaviour of the microcanonical entropy function.

Even so the appearance of a
convex dip in the microcanonical entropy usually signals 
a first order transition,
it may also show up in systems with a continuous transition.
We have shown in this work that its appearance  in those cases (and therefore
also the related appearance of a double-peak structure in
canonical histograms) 
can be traced back to a finite-size effect and that the properties of
the dip, which are markedly different from those associated with a
first order transition, can be understood within a scaling theory 
developed for the description of
continuous phase transitions in the microcanonical ensemble.

\end{document}